\documentclass{article}

\usepackage[preprint]{neurips_2025}

\usepackage[utf8]{inputenc} % allow utf-8 input
\usepackage[T1]{fontenc}    % use 8-bit T1 fonts
\usepackage{hyperref}       % hyperlinks
\usepackage{url}            % simple URL typesetting
\usepackage{booktabs}       % professional-quality tables
\usepackage{amsfonts}       % blackboard math symbols
\usepackage{nicefrac}       % compact symbols for 1/2, etc.
\usepackage{microtype}      % microtypography
\usepackage{xcolor}         % colors
\usepackage{amsmath,amsthm,amssymb}
\usepackage{bm}
\usepackage{graphicx}

\newtheorem{theorem}{Theorem}

\title{A Unified Cortical Circuit Model
with Divisive Normalization and Self-Excitation
for Robust Representation and Memory Maintenance}

\author{%
    Jie SU \\
    Qiyuan Laborotary \\
    \texttt{su.jie@email.cn} \\
    \And
    Weiwei WANG \\
    Beijing Normal University \\
    \texttt{weiwei.wang@mail.bnu.edu.cn} \\
    \And
    Zhaotian GU \\
    Beijing Normal University \\
    \texttt{zhaotiangu@mail.bnu.edu.cn} \\
    \AND
    Dahui WANG \\
    Beijing Normal University \\
    \texttt{wangdh@bnu.edu.cn} \\
    \And
    Tianyi QIAN\thanks{Corresponding author}\\
    Qiyuan Laborotary \\
    \texttt{qiantianyi@qiyuanlab.com} \\
}

\begin{document}
\maketitle

\begin{abstract}

    Robust information representation and its persistent maintenance
    are fundamental for higher cognitive functions.
    Existing models employ distinct neural mechanisms
    to separately address noise-resistant processing or information maintenance,
    yet a unified framework integrating both operations
    remains elusive---a critical gap in understanding cortical computation.
    Here, we introduce a recurrent neural circuit that
    combines divisive normalization with self-excitation
    to achieve both robust encoding and stable retention of normalized inputs.
    Mathematical analysis shows that, for suitable parameter regimes,
    the system forms a continuous attractor with two key properties:
    (1) \emph{input-proportional stabilization} during stimulus presentation;
    and (2) \emph{self-sustained memory states} persisting after stimulus offset.
    We demonstrate the model's versatility in two canonical tasks:
    (a) noise-robust encoding in a random-dot kinematogram (RDK) paradigm;
    and (b) approximate Bayesian belief updating in
    a probabilistic Wisconsin Card Sorting Test (pWCST).
    This work establishes a unified mathematical framework
    that bridges noise suppression, working memory,
    and approximate Bayesian inference within a single cortical microcircuit,
    offering fresh insights into the brain's canonical computation
    and guiding the design of biologically plausible artificial neural architectures.
\end{abstract}

\section{Introduction}

% Biological intelligence has attracted widespread attention
Biological intelligence has garnered widespread attention
due to its ability to efficiently and reliably process information
in complex, dynamic, and uncertain environments.
This capability enables the performance of intricate cognitive tasks,
allowing for flexible and efficient adaptation to changing circumstances.
In order to achieve a reliable understanding of the external environment
and plan subsequent decisions and actions,
the brain must effectively perform at least
two critical cognitive computations:
(i) \emph{Noise-resistant neural coding},
which filters out irrelevant variability
and preserves key signal features for further processing,
and (ii) \emph{Stable maintenance of information},
which ensures that information is held and represented
over time to support memory and planning.
For example, in perceptual tasks
the brain must denoise noisy sensory inputs
to form reliable estimates of motion or contrast
\citep{simoncelli_model_1998,deneve_efficient_2001},
while in working memory and cognitive control
it must sustain internal representations of
stimuli, rules, or values across delays
\citep{wang_synaptic_1999, compte_synaptic_2000, behrens_learning_2007}.

Noise suppression in the brain has been
most extensively studied in sensory cortices
\citep{simoncelli_model_1998, lee_hierarchical_2003,
sawada_divisive_2017, heeger_recurrent_2020,
burg_learning_2021, ernst_dynamic_2021, weiss_modeling_2023}.
While there exists different perspectives,
three major mechanisms contribute to noise suppression.
Divisive normalization accept that neurons' responses
are divided by the activity of a local population
\citep{simoncelli_model_1998,sawada_divisive_2017,heeger_recurrent_2020},
while Bayesian approach consider the neuron population
performing Bayesian inference
\citep{lee_hierarchical_2003,pouget_inference_2003,
knill_bayesian_2004,ma_bayesian_2006}.
Besides this, attractor networks are also considered to be noise resistant
\citep{hopfield_neural_1982,deneve_efficient_2001},
making it a candidate model for noise suppression.

On the other hand, persistent maintenance or working memory
has been modeled primarily using recurrent attractor networks
\citep{wang_probabilistic_2002, wang_neural_2012,
murray_working_2017,bouchacourt_flexible_2019},
or built into network connections through the dynamics of synapses
\citep{compte_synaptic_2000,mongillo_synaptic_2008,morrison_phenomenological_2008,wu_dynamics_2008}.
Discrete attractors can be used for discrimination or classification,
thus also act as decision-making models
\citep{wang_probabilistic_2002,wang_decision_2008},
while continuous attractors are usually considered as
representational manifolds,
acting as relatively sensitive working memory models.

Despite these advances,
noise-resistant encoding and persistent maintenance have remained
largely separate modeling domains:
normalization circuits focus on feedforward gain control
and noise filtering but fade after input removal
\cite{heeger_recurrent_2020},
whereas attractor models preserve persistent activities
yet exhibit limited adaptability,
hindering their extension to novel tasks.
This dichotomy prompts a pivotal question:
by what microcircuit architectures and dynamics does the cerebral cortex
reconcile transient noise filtering with flexible, self-sustained representations?

In this paper, we bridge this gap by proposing
a Recurrent Divisive Normalization (RDN) circuit model:
each excitatory neuron combines its external drive
with self-excitation and then divides by a global inhibitory pool.
We prove that, under proper parameter conditions,
the model not only computes exactly normalized representations
of its inputs (robust to noise and gain changes)
but also forms a continuous attractor
that persistently maintains those representations after input withdrawn.
With some conceptual perceptual (random dot kinematogram)
and cognitive (probabilistic WCST) tasks,
we demonstrated the model's versatility.
By unifying two canonical mechanisms---divisive normalization
and attractor dynamics---this work offers a principled framework
for understanding how the brain performs both
robust filtering and flexible memory within a single circuit motif,
demonstrating the power of this simple model,
and could open up new avenues for bio-inspired artificial intelligence.

\section{A recurrent divisive normalization model}
\label{sec:model}
\begin{figure}[htb!]
    \centering
    \includegraphics{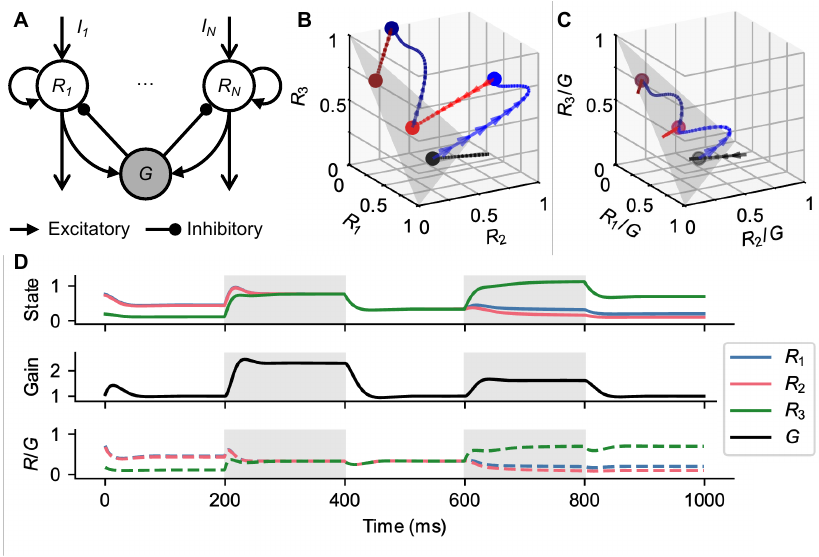}
    \caption{The Recurrent Divisive Normalization (RDN) model and its dynamics.
    \textbf{A}. Schematic of the recurrent divisive normalization circuit;
        Each excitatory unit \(R_i\) combines its external input \(I_i\)
        with a feedback term \(\beta R_i\),
        then divides by \(\eta+G\) (Eq.~\ref{eq:dg}).
        The global pool \(G\) aggregates all \(R_i\) outputs (Eq.~\ref{eq:dr}).
    \textbf{B--C}: Demonstrating the dynamics of a model
        ($N=3$, $\beta=2$, $\eta=1$ and $w_i=1$)
        with and without input.
        The model is randomly initialized, and become steady after a while;
        with a constant input ($I_i = 1$),
        the model changes its state to a new fixed point
        where all $R_i$s are the same,
        and keeps the ratio after input removed.
        Different input ($I_1=0.2, I_2=0.1, I_3=0.7$)
        drives the model to a different fixed point.
    \textbf{B}. Attractor manifold and the trajectory of $R_i$
        with (blue lines) and without (red lines) input.
        The colored dots represents the corresponding fixed points.
    \textbf{C}. Attractor manifold and the trajectory of $R_i/G$
        with (blue lines) and without (red lines) input;
        Fixed points align with/without input become the same.
    \textbf{D}. Dynamics of the model, showing state, gain and the readout
        changing with time during input (shadowed) or not.
        Blue, pink and green represents the 3 excitatory neurons $R_i$,
        black represent the inhibitory neuron $G$.
    }
    \label{fig:model}
\end{figure}

We present a firing-rate model that accomplish
divisive normalization and self-excitatory recurrent connection.
Consider \(N\) excitatory units \(R_i\)
coupled to a single inhibitory pool \(G\) (Fig.~\ref{fig:model}A),
where the dynamics of the network
is defined by ordinary differential equations:
\begin{align}
    \tau_R \frac{\mathrm{d}R_i}{\mathrm{d} t}
        &= -R_i + \frac{\beta R_i + I_i}{\eta+G},
        \quad i=1, \cdots, N \label{eq:dr} \\
    \tau_G \frac{\mathrm{d}G}{\mathrm{d}t}
        &= -G+\sum_{i=1}^N w_i R_i.
        \label{eq:dg}
\end{align}
Key parameters include:
\(I_i \ge 0\) (external input to unit \(i\)),
\(\beta \ge 0\) (self-excitation strength),
\(\eta > 0\) (semi-saturation constant),
\(w_j > 0\) (excitatory-to-inhibitory weights),
and \(\tau_R, \tau_G > 0\) (time constants).
% We denote $\bm{R} = (R_1, \dots, R_N)^\top \in \mathbb{R}^N$
% be the state vector of the model.

When \(\beta=0\), the model reduces to
the classical divisive normalization
\citep{heeger_normalization_1992,simoncelli_model_1998,
keung_divisive_2020},
and \(\beta=1\) lead to a model similar to some well-known
recurrent divisive normalization models
\cite{heeger_oscillatory_2019, heeger_recurrent_2020,NIPS24_unconditional},
making our model as a generalization of divisive normalization model.
Despite its structural simplicity,
our model exhibits remarkable dynamical properties (Fig.~\ref{fig:model}).
Under appropriate parameter configurations,
it forms a continuous attractor networks capable of
executing diverse neurocomputational functions,
including noise-resistant information processing,
persistent memory maintenance, and dynamic learning adaptation.
% We will start with a theoretical analysis of the model properties.

\section{Model analysis}

\subsection{Steady-state solutions}
\label{sec:steady}

Setting \(\dot R_i = \dot G = 0\).
From Eq.~\eqref{eq:dr}, we have
\begin{equation*}
    R_i^*(\eta + G^*) = \beta R_i^* + I_i,
    \quad
    R_i^* = \frac{I_i}{\eta - \beta + G^*}.
\end{equation*}
Substituting into Eq.~\eqref{eq:dg}, let
\begin{equation*}
    T = \sum_{j=1}^N w_j I_j,
\end{equation*}
thus
\begin{align*}
    G^* = \sum_{j=1}^N w_j R_j = \frac{T}{\eta - \beta + G^*},
    \quad
    G^*(\eta - \beta + G^*) = T.
\end{align*}
We obtain a quadratic equation of $G^*$:
\begin{equation}
    (G^*)^2 + (\eta - \beta) G^* - T = 0.
    \label{eq:gfixed}
\end{equation}
The solution is given by
\begin{equation*}
    G^* = \frac{
        -(\eta - \beta) \pm \sqrt{(\eta - \beta)^2 + 4T}
    }{2}.
\end{equation*}
For $I_i \ge 0$ and $T > 0$, the discriminant
\(\Delta = (\eta - \beta)^2 + 4T > 0\),
and it is easy to obtain that Eq.~\eqref{eq:gfixed}
has exactly one root satisfying $G^* > 0$:
\begin{equation}
    G^* = \frac{
        -(\eta - \beta) + \sqrt{(\eta - \beta)^2 + 4T}
    }{2},
    \label{eq:gpos}
\end{equation}
with $\eta - \beta + G^* > 0$ also be satisfied,
implying $R_i^* \ge 0$.
Therefore, the RDN model has a unique physiologically meaningful
steady-state solution \((\bm{R}^*, G^*) \in \mathbb{R}^{N+1}\).

To analyze the stability of the model,
We apply the indirect method of Lyapunov at the fixed point
\citep{strogatz_nonlinear_2018}.
Linearize $x = (\delta R_1, \dots, \delta R_N, \delta G)$.
The Jacobian is
\begin{equation}
    \bm{J} =
    \begin{bmatrix}
        a \bm{I}_N & \bm{b} \\
        \bm{c}^\top & -1/\tau_G
    \end{bmatrix},
\end{equation}
where $\bm{I}_N$ is an $N$-th order identity matrix,
$\bm{b} = (b_1, \dots, b_N)^\top$
and $\bm{c} = (c_1, \dots, c_N)^\top$,
\begin{equation*}
    a = \frac{
        -1 + \beta/(\eta + G^*)
    }{\tau_R},
    \quad
    b_i = - \frac{\beta R_i^* + I_i}{\tau_R (\eta + G^*)^2},
    \quad
    c_j = \frac{w_j}{\tau_G}.
\end{equation*}
Eigendecomposition of the Jacobian matrix
$\bm{J}$ reveals fixed-point stability
\citep{strogatz_nonlinear_2018}:
\begin{enumerate}
    \item Eigenvalues with negative real parts indicate
    contracting dynamics in corresponding state-space dimensions
    \item Positive real parts correspond to expanding dynamics
    \item Zero real parts reflect marginal stability.
\end{enumerate}
% for the eigenvalue $\lambda$ with a negative real part,
% the dynamics are contracting in the corresponding dimension of state space,
% while that with a positive real part means expanding dynamics
% and that of zero real part refers to marginal stability
Stable attractors might be useful for integrating noisy information,
or for maintaining state or memory.

From Theorem~\ref{eigenval} (see Appendix),
we know that the eigenvalues of $\bm{J}$ are
\begin{align}
    \lambda_i &= a = \frac{
        -1 + \beta/(\eta + G^*)
    }{\tau_R},\quad (i=1, \dots, N-1), \label{eq:lambda-a}\\
    \lambda_\pm &= \frac{
        (a - 1/\tau_G) \pm \sqrt{
            % (a + 1/\tau_G)^2 + 4 \sum_j b_j c_j
            (a + 1/\tau_G)^2 - 4 G^*/(\eta + G^*)/(\tau_R\tau_G)
        }
    }{2} \label{eq:lambda-pm}.
\end{align}

To analyse the stability of the fixed point analytically
is intractable through the above eigenvalues.
With simplified parameters
$\tau_R = \tau_G = \tau$ and $w_i = 1$ for all $i=1, \dots, N$,
where the inhibitory pool receives the neurons' inputs equally
and have the same time constant with excitatory units,
we next analyze the simplified model with the inputs
$I_i$ removed from and exerted to it.

\subsection{Stability of the model without inputs}
\label{sec:noinput}

With the above settings and $I_i \equiv 0$,
the dynamics of the model is simplified as
\begin{align}
    \tau \frac{\mathrm{d}R_i}{\mathrm{d} t}
        &= -R_i + \frac{\beta R_i}{\eta+G},
        \quad i = 1, \cdots, N \label{eq:dr0} \\
    \tau \frac{\mathrm{d}G}{\mathrm{d}t}
        &= -G + \sum_{j=1}^N{R_j}, \label{eq:dg0}
\end{align}
Eq.~\eqref{eq:gfixed} become
\begin{equation*}
    G^* (\eta - \beta + G^*) = 0.
\end{equation*}
This solves the fixed point $G^* = R_1^* = \cdots = R_N^* = 0$,
which is a trivial fixed point,
and $G^*=\sum_{i=1}^N{R_i^*} = \beta - \eta$,
which forms an $N-1$ dimensional continuous attractor manifold
(hyperplane) in the $\bm{R}$ subspace
(Fig.~\ref{fig:model}B).

% We then apply the indirect method of Lyapunov
% at these fixed points for this system
% \citep{strogatz_nonlinear_2018},
% and have the Jacobian matrix defining
% the linearized system near the fixed points
% \begin{equation}
%     \mathbf{J} = \frac{1}{\tau}
%     \begin{bmatrix}
%         \left( -1+\frac{\beta}{\eta+G^*} \right)\mathbf{E}_N
%         & -\frac{\beta \mathbf{R}^*}{(\eta+G^*)^2} \\
%         \mathbf{1}^T & -1\\
%     \end{bmatrix}
%     \label{eq:jacobian}
% \end{equation}
% where $\mathbf{R}^*=[R^*_1, R^*_2, \cdots, R^*_N]^T$.

\begin{figure}[htb!]
    \centering
    \includegraphics{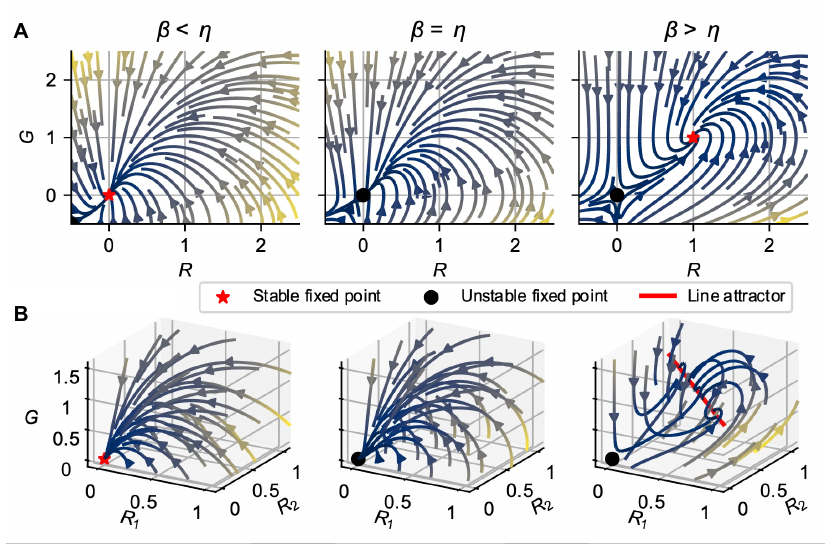}
    \caption{Phase portrait of the model before and after
        bifurcation (at $\beta=\eta$).
    \textbf{A}. Phase portrait of the simplified model
        Eqs.~(\ref{eq:dr0}--\ref{eq:dg0}) with one excitatory node ($N=1$),
        with $\beta>\eta$, there exist a non-zero stable fixed point $(1,1)$,
        which makes our RDN model fundamentally different
        from the canonical divisive normalization model,
        enabling persistent memory.
    \textbf{B}. Phase portrait of the simplified RDN model
        with two excitatory node ($N=2$),
        showing the same bifurcation properties with the 1-Node model (A).
        Number of excitatory neurons can easily extended
        to fit different kind of tasks,
        making the RDN model much more useful compared to other models.
    }
    \label{fig:bifur}
\end{figure}

For the trivial fixed point $G^*=R_1^*=\cdots=R_N^*=0$,
the eigenvalues simplified as
\begin{equation}
    \begin{aligned}
        \lambda_{1, \cdots, N} &= \frac{\beta - \eta}{\tau \eta},
        \lambda_{N+1} &= -\frac{1}{\tau}.
    \end{aligned}
\end{equation}
Therefore, the trivial fixed point will be stable
and the system will converge
if and only if $\beta < \eta$ (Fig.~\ref{fig:bifur}, left panel).

For the continuous attractor
$G^* = \sum_{i=1}^N{R^*_i} = \beta-\eta$,
eigenvalues $\lambda_{1, \cdots, N-1} = 0$, and
% we have
% \begin{equation}
%     \mathbf{J} = \frac{1}{\tau}
%     \begin{bmatrix}
%         \mathbf{0} & -\frac{\mathbf{R^*}}{\beta} \\
%         \mathbf{1}^T & -1 \\
%     \end{bmatrix}.
% \end{equation}
% This matrix possesses the eigenvalue
% $\lambda_{1, \cdots, N-1} = 0$, and
% which refers to the \emph{marginal stability} of the continuous attractor.
% The other eigenvalues are
\begin{equation}
    \lambda_{N, N+1} = \frac{-1 \pm \sqrt{4\eta/\beta - 3}}{2\tau}.
\end{equation}
Therefore, the continuous attractor will be stable
and the system will converge to the continuous attractor
$G^*=\sum_{i=1}^N{R^*_i} = \beta-\eta$
if and only if $\beta > \eta$ (Fig.~\ref{fig:bifur}, right panel).

Taken together, the dynamical system governed by
Eqs.~(\ref{eq:dr0}--\ref{eq:dg0})
exhibits fundamentally distinct attractor structures in phase space
and markedly differentiated dynamical characteristics
when the parameters $\beta, \eta$ crossing critical point,
yields a transcritical bifurcation in the state space
(See Fig.~\ref{fig:bifur} for demonstrations,
see also Fig.~\ref{fig:model}B\&D the detailed dynamics
with 3 excitatory node ($N=3$),
and the attractor manifold $G^*=\sum_{i=1}^{3} R_i = 1$).

Next we discuss the dynamics when the non-zero inputs
are \emph{withdrawn} from the system,
which is crucial for the working memory capability of the model.
Denote $S = \sum_{i=1}^N R_i, \rho_i = R_i/S$,
which satisfies $\sum_{i=1}^N \rho_i = 1$, we can deduce that
\(
\tau \mathrm{d} \rho_i/\mathrm{d} t = 0
\)
(see Eq.~\eqref{eq:propto}, Appendix).

% \begin{equation}
%     \begin{aligned}
%     \tau \frac{\mathrm{d} \rho_i}{\mathrm{d} t}
%     &= \frac{\tau}{S} \frac{\mathrm{d}R_i}{\mathrm{d}t}
%         - \frac{\tau R_i}{S^2} \frac{\mathrm{d}S}{\mathrm{d}t} \\
%     &= \frac{\tau}{S}\frac{1}{\tau}\left(-R_i+\frac{\beta R_i}{\eta+G}\right)
%         - \frac{\tau R_i}{S^2}\sum_{j=1}^N \frac{\mathrm{d}R_j}{\mathrm{d}t} \\
%     &= \frac{1}{S}\left(-R_i+\frac{\beta R_i}{\eta+G}\right)
%         - \frac{\tau R_i}{S^2}\frac{1}{\tau}\sum_{j=1}^N\left(
%             -R_j+\frac{\beta R_j}{\eta+G}
%         \right) \\
%     &= \frac{1}{S}\left(-R_i+\frac{\beta R_i}{\eta+G}\right)
%         - \frac{R_i}{S^2}\left(
%             -S+\frac{\beta S}{\eta+G}
%         \right) \\
%     &= 0
%     \label{eq:propto}
%     \end{aligned}
% \end{equation}

When the system converges to the continuous attractor
$G^*=\sum_{i=1}^N R^*_i = \beta-\eta$ eventually,
the firing rate $R_i$ keeps constant at $\rho_i (\beta-\eta)$,
where $\rho_i$ relies on the state before the inputs are withdrawn.
In other words, % the system stores the memory of previous inputs
the system maintains the last input-normalized pattern indefinitely,
implementing working memory (Fig.~\ref{fig:model}B\&D).

\subsection{Stability of the model with inputs}
\label{sec:input}
With inputs $I_i$ exerted to the system, we have
\begin{align}
    \tau \frac{\mathrm{d}R_i}{\mathrm{d} t}
        &= -R_i + \frac{\beta R_i + I_i}{\eta+G},
        \quad i = 1, \cdots, N \label{eq:drI} \\
    \tau \frac{\mathrm{d}G}{\mathrm{d}t}
        &= -G + \sum_{j=1}^N{R_j} \label{eq:dgI},
\end{align}
% Let
% \begin{equation}
%     \begin{aligned}
%         -R_i^*+\frac{I_i+\beta R_i^*}{\eta+G^*} &= 0, i=1, 2, \cdots, N \\
%         -G^*+\sum_{i=1}^n R_i^* &= 0
%     \end{aligned}
% \end{equation}
Solve as like in Eq.~\eqref{eq:gpos}, we have the fixed point
\begin{align}
    R_i^* &= \frac{2 I_i}{
        (\eta-\beta) + \sqrt{(\eta-\beta)^2 + 4T}
        }, \\
    G^* &= \frac{(\beta-\eta) + \sqrt{(\eta-\beta)^2 + 4T}}{2}.
\end{align}
which reveals that the system shifts toward a new fixed point
while receiving new inputs. Since
\begin{equation}
    \begin{aligned}
        \frac{R_i}{G} &= \frac{2 I_i}{
            \sqrt{(\eta-\beta)^2 + 4 T} + (\eta - \beta)
        } \cdot \frac{2}{
            \sqrt{(\eta-\beta)^2 + 4 T} - (\eta - \beta)
        } \\
        &= \frac{4 I_i}{(\eta-\beta)^2 + 4 T - (\eta-\beta)^2}
        = \frac{I_i}{T} = \frac{I_i}{\sum_{j=1}^n{I_j}},
    \end{aligned}
\end{equation}
therefore the system eventually evolves to a fixed point
proportional to the input,
and one can define a readout rule from the model such that
$O_i = R_i / G$, which essentially normalize the inputs
(Fig.~\ref{fig:model}C\&D).

Given $I_i > 0$, Eq.~\eqref{eq:lambda-pm}
demonstrates that all eigenvalues satisfy $\lambda_i < 0$ 
under the condition $\beta < \eta + G^*$.
This inequality is inherently guaranteed by Eq.~\eqref{eq:gpos}.
Thus the unique steady-state forms an attractor
(Theorem~\ref{stability}, see Appendix).
One can deduce that 
the firing rate $R_i$ of each neuron is proportional to
the strength of input $I_i$ if the system has converged to the fixed point,
following the procedures described in
Eq.~(\ref{eq:propto1}--\ref{eq:propto2}).
It is also worth noting that with the inputs withdrawn,
the firing rate $R_i$ will converge to a steady state
$R_i = \rho_i(\eta - \beta)$,
where $\rho_i$ relies on the state before the inputs were withdrawn.
Therefore, the system stores the memory of the latest inputs,
which might be attributed to many cognitive tasks.

\section{Application}

\subsection{Perceptual denosing}
% \subsection{Perceptual Decision-Making (RDK Task)}

Perceptual denoising is critical for sensory processing,
enabling the brain to filter out noise while preserving accurate representations of sensory inputs.
To evaluate the denoising and maintenance capabilities
of our Recurrent Divisive Normalization (RDN) model,
we employed the Random Dot Kinematogram (RDK) task,
a well-established paradigm for studying sensory perception
under noisy conditions \citep{williams_coherent_1984}.

\begin{figure}[htb!]
    \centering
    \includegraphics{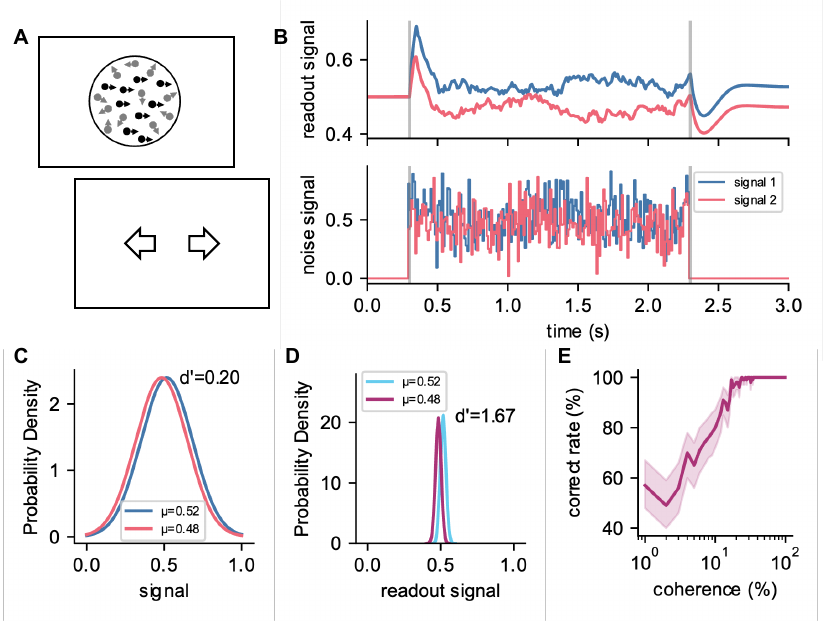}
    \caption{
        The Recurrent Divisive Normalization model exhibits noise reduction,
        making it useful in sensory perception task (RDK).
    \textbf{A}. Paradigm of the RDK task.
        A large number of randomly moving dots were presented
        with a dominant direction (left or right).
        The percentage of coherently moving dots (coherence, 0-100\%)
        controls task difficulty.
        Lower coherence trials are more difficult with more noises.
        % Simulated RDK task: time courses of \(R_L,R_R\) under low coherence.
	% The model filters noise and integrates evidence,
	% with \(O_{L,R}=R_{L,R}/(R_L+R_R)\) converging to correct choice.
    \textbf{B}. The model filters noise and gives denoised signals.
        Two inputs, ``signal 1'' and ``signal 2'' (lower panel),
        were used to simulate motion strength of the RDK task,
        corresponding to leftward and rightward motion of 10\% coherence level.
        These signals are input to
        The stimuli was given at 300 ms and last for 2000 ms.
        Readout represents the model's state,
        showing the processed signals (upper panel). 
    \textbf{C}. Probability density of the input signal,
        due to noise, the two signals are difficult to distinguish ($d'=0.20$).
    \textbf{D}. Distribution of the ``readout'' signal from the model,
        showing much reduced noises.
        The processed signals can be better distinguished
        compared to the input signal($d'=1.67$).
        % The distribution is estimated from a stable signal
        % readout 1 s after the input of a 30-s noise signal.
    \textbf{E}. Relationship between coherence of the RDK task
        and the accuracy of a simple discriminator
        using the processed signal (denoised representations) for choice.
        % The accuracy is assessed using a 1-s input of random signals,
        % with the readout state judged 1 s after the input is removed.
        Each coherence level is sampled 100 times.
        Error bars represent the 95\% confidence intervals (CIs).
    }
    \label{fig:rdk}
\end{figure}

In the RDK task, a number of dots are randomly,
with a subset moving coherently in a dominant direction
(Fig.~\ref{fig:rdk}A).
Task difficulty is modulated by coherence,
defined as the percentage of dots moving towards the dominant direction.
For this study, we simulated a 2-Alternative Foice Choice (2AFC)
motion discrimination task (left vs. right)
using two independent Gaussian distributions
with identical variance ($\sigma^2$)
but distinct means ($\mu_1, \mu_2$, Fig.~\ref{fig:rdk}C),
where the separation between means
was controlled by the coherence parameter:
\begin{equation}
    \text{coherence} = \frac{|\mu_1 - \mu_2|}{\sigma}
    \label{eq:coherence}
\end{equation}

We use RDN model with $N=2$ for this task,
where \(R_L, R_R\) encode evidence for left/right motion respectively.
The model parameters were set to $\tau=50 \text{ms}$, $\beta=2$ and $\eta=1$,
with a numerical simulation time step of $dt = 0.1 \text{ms}$.
Random input signals were sampled at 100 Hz
and simulated using zero-order hold interpolation.
In our simulations,
input signals represented motion strength
with means of 0.52 (signal 1) and 0.48 (signal 2)
and a shared variance of 0.17 (Fig.~\ref{fig:rdk}B, lower panel).
Stimuli were presented for 2 s (onset: 300 ms).
The model's readout signal (Fig.~\ref{fig:rdk}B, upper panel)
exhibited noise reduction significantly compared to the raw inputs.
Probability density analyses (Fig.~\ref{fig:rdk}C\&D)
revealed that the original signals were
indistinguishable due to noise ($d' = 0.20$),
whereas the readout signals showed clear separation ($d' = 1.67$).
D-prime ($d'$) is a measure of sensitivity in signal detection theory, calculated as:
\begin{equation}
    d' = \frac{\mu_S - \mu_N}{\sqrt{\frac{\sigma_S^2 + \sigma_N^2}{2}}},
    \label{eq:dprime}
\end{equation}
where $\mu_S$ and $\mu_N$ are the means of the signal and noise,
$\sigma_S^2$ and $\sigma_N^2$ are their variances.

The readout maintained stable mean values with reduced variance,
demonstrating robust noise suppression.
Moreover, the model sustained its representation
after stimulus offset (Fig.~\ref{fig:rdk}B),
highlighting its capacity of persistence memory.

To quantify representational accuracy across coherence levels,
we simulated 100 coherence levels (1-100\%).
The readout state was assessed 1 s post-stimulus,
with correctness determined by alignment
to the ground-truth distribution (100 samples/condition).
Accuracy increased monotonically with coherence,
reaching 100\% at ~20\% coherence (Fig.\ref{fig:rdk}E),
consistent with primate neurophysiological data
(\cite{britten_analysis_1992, wang_probabilistic_2002},
e.g., MT/LIP activity).

In summary, the RDN model achieves effective noise filtering
and stable representation maintenance through divisive normalization
and $\beta$-mediated feedback.
Its performance aligns with psychophysical curves observed
in primate dorsal stream areas,
underscoring its utility for sensory decision-making under uncertainty.

\subsection{Probabilistic inference}
% \subsection{Probabilistic rule learning (pWCST)}

\begin{figure}[htb!]
    \centering
    \includegraphics[width=\linewidth]{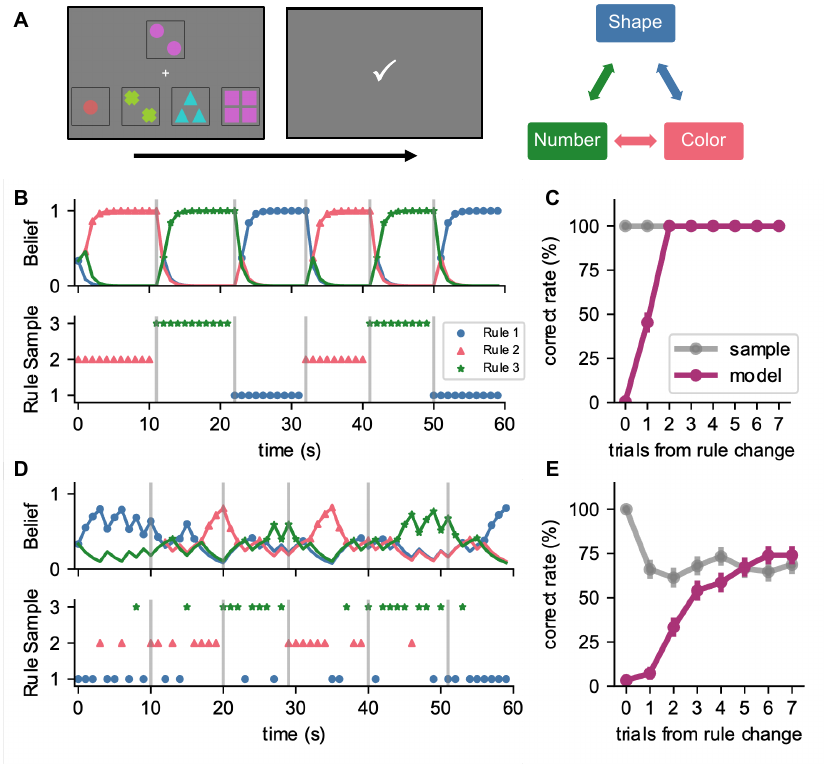}
    \caption{
        The Recurrent Divisive Normalization Model performs
        probabilistic inference in classical and probabilistic WCST task.
    \textbf{A}. The WCST task paradigm and its flexible changing structure:
        A reference card and some candidate cards are presented.
        Choices are evaluated based on matching
        shape (Rule 1), number (Rule 2), or color (Rule 3),
        with feedback indicating correctness or error
        according to the current rule,
        which should be used to learn the randomly switching rule.
    \textbf{B}. Rule design of a classical WCST task
        and the learned belief by the RDN model.
        Each block in the classic WCST employs a single,
        fixed rule for discrimination, comprising 9, 10, or 11 trials
        with a 1-second interval between trials.
        Different blocks are separated by gray vertical lines.
        Rule Sample refers to the rule used for discrimination,
        (Rule 1: blue circle, Rule 2: pink triangle, Rule 3: green pentagram).
        the model selects the rule with the highest belief
        as the basis for its choice in each trial
        (marked with the same symbols) and update belief based on feedback.
    \textbf{C}. Correct rate after change point show that
        the tested RDN model makes optimal rule learning and switching,
        i.e., win-stay, loss-shift.
        Gray circles represent simulated sampling data,
        purple circles indicate the model's accuracy.
        Error bars denote the standard error of the mean (SEM).
    \textbf{D}. Rule design of a probabilistic WCST (pWCST) task
        and model performance in the task.
        Each block employs probabilistic rules for discrimination,
        with a dominant rule having a probability of 0.7,
        making it impossible to relay on single feedback.
        The RDN model can learn and track the rule switch very well.
        Notations are consistent with B.
    \textbf{E}. Correct rate after probability change point show that
        the tested RDN model works well on the pWCST task.
        Note that to ensure accurate switch timing,
        the first and last trial of each block are fixed to the dominant rule.
        Notations are consistent with C.
    }
    \label{fig:wcst}
\end{figure}

Probabilistic inference is fundamental to adaptive behavior,
particularly in tasks requiring flexible rule learning and switching
\citep{behrens_learning_2007,behrens_associative_2008,boorman_how_2009}.
Cognitive flexibility are often measured using
tasks like the Wisconsin Card Sorting Test (WCST)
\citep{stemme_neurodynamics_2007},
which can be modeled as probabilistic inference
\citep{dalessandro_bayesian_2020}
or flexible gating \citep{liu_flexible_2024}.
To evaluate the RDN model's capacity for belief updating,
we examined its performance in both the classical
and probabilistic versions of WCST.
The classical WCST task assesses executive functions,
including cognitive flexibility and attentional control.
Participants match candidate cards to a reference card
based on a latent rule (e.g., shape number or color; Fig.~\ref{fig:wcst}A).
Feedback indicates correctness, and rules switch pseudorandomly,
demanding rapid adaptation.

In our simulations (Fig.~\ref{fig:wcst}B),
the RDN model comprised three excitatory units,
encoding the belief of the corresponding rules.
We use $\tau=50 \text{ms}$, $\beta=2$ and $\eta=1$ for this task,
and $dt = 0.1 \text{ms}$ for numerical simulations.
Feedback inputs were set to 1 for unit corresponding to chosen rule
and 0 for other units at correct trials,
otherwise 0 for the chosen rule and 0.5 for others.
Feedback inputs lasted for 50 ms per trial.
Rules persisted for 9--11 trials before switching.
The model achieved rapid rule-switching detection
within two trials (Fig.~\ref{fig:wcst}B\&C),
consistent with theoretically optimal strategy bounds.
Analysis of 150 rule blocks (Fig.~\ref{fig:wcst}C) confirmed this behavior,
demonstrating the model's ability
to leverage feedback for exact belief updating.

% In a probabilistic WCST, each unit \(R_i\) represents belief in rule \(i\).
% Feedback-driven inputs \(I_i\) update beliefs,
% and normalization ensures \(\sum_i R_i=1\).
% The attractor dynamics maintain beliefs through inter-trial intervals.
% Preliminary simulations align with human adaptation rates
% \citep{dalessandro_bayesian_2020}.

The probabilistic WCST task introduces greater complexity:
feedback follows the dominant rule probabilistically (70\%),
occasionally adheres to incorrect rules (30\%).
This requires integration of historical reward to mitigate stochastic feedback,
rather than adopt the ``win-stay, loss-shift'' strategy.
With $\tau$ adjusted to 50 ms (other parameters unchanged),
the model maintained robust performance (Fig.~\ref{fig:wcst}D\&E).
With a 150-block experiment,
the accuracy from rule change point stabilized
near the probability of the dominant rule(0.7),
reflecting effective longer-term reward integration despite interference.

The RDN model achieves rapid rule switching in deterministic contexts
and sustains performance under probabilistic feedback,
mirroring human adaptive learning.
Its success in both WCST variants underscores its utility
for modeling approximate Bayesian inference in dynamic environments
\citep{behrens_learning_2007}.

\section{Discussion and conclusion}

% Divisive normalization is recognized as a canonical neural computation
% across brain systems.

Our study presents a unified cortical circuit model that
integrates divisive normalization with self-excitation
to achieve both robust sensory processing and stable memory maintenance.
This work bridges two fundamental but traditionally separate lines of research:
normalization as a canonical cortical operation for noise-resistant encoding
\citep{carandini_normalization_2011},
and attractor dynamics for persistent information storage
\citep{compte_synaptic_2000, wang_probabilistic_2002}.
The model's mathematical framework demonstrates
how these mechanisms coexist within a single microcircuit,
offering a parsimonious alternative to modular
architectures requiring specialized subsystems.
Functionally, it replicates both noise-robust sensory encoding
\citep{britten_analysis_1992}
and cognitive flexibility in rule-switching tasks
\citep{liu_flexible_2024},
suggesting common computational principles
may underlie diverse neural functions.

The model resolves a critical theoretical gap.
While divisive normalization has been formalized
as a fundamental nonlinear operation
\citep{kouh_canonical_2008},
and attractor networks are well-established for memory maintenance
\citep{constantinidis_neuroscience_2016},
their integration remained unexplored.
Our framework reveals that normalization not only suppresses noise
but also stabilizes attractor dynamics against
input fluctuations---a prediction testable through
combined electrophysiology and perturbation experiments.
% This aligns with emerging views of cortex as implementing
% ``canonical computations'' \citep{heeger_theory_2017},
% while uniquely demonstrating their synergy in a minimal circuit.
While aligning with the emerging concept of cortical
``canonical computations'' \citep{heeger_theory_2017},
our work uniquely demonstrates how these computations
synergize within a minimal circuit architecture,
while may represent a plausible neural-circuit implementation
of the Bayesian brain hypothesis
\citep{knill_bayesian_2004,friston_free-energy_2009}.

Several limitations highlight future directions.
First, while the model captures core phenomena,
predictions like persistent normalization during memory delays
await experimental validation---a challenge shared
by many theoretical studies
\citep{sreenivasan_revisiting_2014}.
Second, rate-based dynamics simplify biological details
such as spiking neurons and dendritic computations \citep{larkum_guide_2022}.
Third, pre-tuned weights omit developmental plasticity,
though this simplification is common in foundational work
\citep{stemme_neurodynamics_2007}.

Future research should pursue:
Experimental tests of neural signatures
(e.g., normalized firing in prefrontal memory tasks,
brain region specific parameters emphasis different functions);
Introduce nonlinear activation functions to enable the model
to exhibit richer dynamic properties and perform more cognitive tasks;
Spiking implementations to assess temporal precision;
Plasticity mechanisms for self-organized circuit tuning.
% \begin{enumerate}
%     \item Experimental tests of neural signatures
%         (e.g., normalized firing in prefrontal memory tasks);
%     \item Spiking implementations to assess temporal precision;
%     \item Plasticity mechanisms for self-organized circuit tuning.
% \end{enumerate}

By unifying noise suppression and memory maintenance in a single architecture,
this work challenges modular brain views and offers new principles for bio-inspired AI.
The demonstrated synergy of canonical computations
opens avenues to understand cortical efficiency and design adaptive neural systems.

\begin{ack}

\section*{Author contributions}
J.S. designed the architecture,
J.S., W.W. designed and performed the experiments,
J.S., Z.G. analyzed the model.
All authors designed the study,
took part in discussions,
interpreted the results,
and wrote the paper.

% \section*{Acknowledgments}
% This work was supported by xxx.
% The authors would like to acknowledge XXX for xxx

% Use unnumbered first level headings for the acknowledgments. All acknowledgments

% Do {\bf not} include this section in the anonymized submission, only in the final paper. You can use the \texttt{ack} environment provided in the style file to automatically hide this section in the anonymized submission.
\end{ack}

% \section*{References}

% References follow the acknowledgments in the camera-ready paper. Use unnumbered first-level heading for
% the references. Any choice of citation style is acceptable as long as you are
% consistent. It is permissible to reduce the font size to \verb+small+ (9 point)
% when listing the references.
% Note that the Reference section does not count towards the page limit.
% \medskip

% \nocite{*}
\bibliographystyle{unsrt}
% \bibliographystyle{apalike}
% \bibliography{refs}

{
\small
\bibliography{refs}
}

%%%%%%%%%%%%%%%%%%%%%%%%%%%%%%%%%%%%%%%%%%%%%%%%%%%%%%%%%%%%

\appendix
\clearpage

\section{Technical Appendices and Supplementary Material}
% Technical appendices with additional results, figures, graphs and proofs
% may be submitted with the paper submission before the full submission deadline (see above),
% or as a separate PDF in the ZIP file below before the supplementary material deadline.
% There is no page limit for the technical appendices.

\subsection{Eigenvalues of a type of arrowhead matrix}
\begin{theorem}
    \label{eigenval}
Consider the following $N+1$ dimensional arrowhead matrix:
\begin{equation}
    \bm{M}=
    \begin{bmatrix}
        a\bm{I}_N & \bm{v} \\
        \bm{w}^\top & b
    \end{bmatrix},
\end{equation}
where $a, b \in R$ are scalars,
$\bm{I}_N$ is an $N$-th order identity matrix,
and $\bm{v}, \bm{w}$ are $N$-dimensional vectors,
$\bm{w} \ne \bm{0}$.
The eigenvalues of matrix $\bm{M}$ are given explicitly by:
\begin{enumerate}
    \item $\lambda = a$ with multiplicity $N-1$;
    \item Two distinct eigenvalues:
\begin{equation}
    \lambda_\pm = \frac{
        (a+b) \pm \sqrt{(a-b)^2 + 4 \bm{w}^\top \bm{v} }
    }{2}.
\end{equation}
\end{enumerate}
which are real if \((a-b)^2 + 4 \bm{w}^\top \bm{v} \ge 0\),
and form a complex conjugate pair otherwise.
% which satisfies the equation
% \begin{equation*}
%     (\lambda-a)(\lambda-b) - \bm{w}^\top\bm{v} = 0.
% \end{equation*}
\end{theorem}

\begin{proof}
Define a $(N-1)$-dimensional subspace
\begin{equation}
    \mathcal{V}_\perp = \{\bm{u} \in \mathbb{R}^N \mid \bm{w}^\top\bm{u} = 0\}
\end{equation}

For any $\bm{u} \in \mathcal{V}_\perp$,
% one can verify that vector $[\bm{u}, 0]$ satisfies the eigen equation
the vector $[\bm{u}, 0]^\top$ satisfies
\begin{equation*}
    \bm{M}
    \begin{bmatrix}
        \bm{u} \\
        0
    \end{bmatrix}
    =
    \begin{bmatrix}
        a \bm{I}_N \bm{u} + \bm{v}\cdot 0 \\
        \bm{w}^\top\bm{u} + b \cdot 0
    \end{bmatrix}
    = a %% multiply
    \begin{bmatrix}
        \bm{u} \\
        0
    \end{bmatrix}
\end{equation*}
which refers to $(N-1)$-fold eigenvalues corresponding to
the eigenvector $[\bm{u}, 0]^\top$.

Let the remaining eigenvectors have the form
$[m \bm{v}, n]^T$,
with $m, n \in \mathbb{R}$
and substitute it into the eigen equation, we have
\begin{equation*}
\begin{cases}
    m a \bm{v} + n \bm{v} = \lambda m \bm{v} \\
    m \bm{w}^\top \bm{v} + b n = \lambda n.
\end{cases}
\end{equation*}

Simplify to obtain
\begin{equation*}
    \begin{aligned}
    & n = m(\lambda-a), \\
    & m (\bm{w}^\top\bm{v}) + n (b-\lambda) = 0
    \end{aligned}
\end{equation*}
and after eliminating $n$, we obtain
\begin{equation*}
    m (\bm{w}^\top\bm{v}) + m(\lambda-a)(b-\lambda)=0
\end{equation*}
which can be decompsed into a trivial equation $m=0$ and equation
\begin{equation}
    (\lambda-a)(\lambda-b) - \bm{w}^\top\bm{v} = 0
\end{equation}

Solving this quadratic equation explicitly,
we obtain the remaining two eigenvalues:
\begin{equation*}
    \lambda_{\pm} = \frac{(a+b) \pm \sqrt{(a-b)^2 + 4 \bm{w}^\top\bm{v}}}{2}.
\end{equation*}

Note that the discriminant
\begin{equation*}
    \Delta = (a - b)^2 + 4 \bm{w}^\top \bm{v}
\end{equation*}
may be negative.
In that case, the two nontrivial eigenvalues
\( \lambda_{\pm} \) form a complex conjugate pair.

Therefore, the eigenvalues of \(\bm{M}\) are fully characterized by:
\begin{equation*}
    \{\underbrace{a, \dots, a}_{N-1}, \lambda_{+}, \lambda_{-}\}.
\end{equation*}

This completes the proof.
\end{proof}

\subsection{Stability of the simplified model with inputs}

\begin{theorem}
    \label{stability}
    Given dynamic system defined with Eqs.~(\ref{eq:drI}--\ref{eq:dgI}),
    For any parameters $\tau > 0$, $\beta > 0$, $\eta > 0$
    and input $I_i > 0$,
    the unique steady-state $(\bm{R^*}, G^*)$ is an attractor.
\end{theorem}

\begin{proof}
From Eq.~(\ref{eq:lambda-a}--\ref{eq:lambda-pm}),
the eigenvalues of the system is given by
\begin{align}
    \lambda_i &= a = \frac{
        -1 + \beta/(\eta + G^*)
    }{\tau},\quad (i=1, \dots, N-1), \label{eq:lmd-a}\\
    \lambda_+ &= \frac{
        (a - 1/\tau) + \sqrt{
            (a + 1/\tau)^2 - 4 G^*/(\eta + G^*)/\tau^2
        }
    }{2}, \label{eq:lmd-pos} \\
    \lambda_- &= \frac{
        (a - 1/\tau) - \sqrt{
            (a + 1/\tau)^2 - 4 G^*/(\eta + G^*)/\tau^2
        }
    }{2}, \label{eq:lmd-neg}
\end{align}

It is obviously, $\lambda_i = a < 0$ and $\Re(\lambda_-) < 0$
if $\beta < \eta + G^*$, while from Eq.~\eqref{eq:gpos}, we have
\begin{equation}
    \begin{aligned}
        G^* + \eta - \beta &= \frac{
            -(\eta - \beta) + \sqrt{(\eta - \beta)^2 + 4T}
        }{2} + \eta - \beta \\
        &= \frac{
            (\eta - \beta) + \sqrt{(\eta - \beta)^2 + 4T}
        }{2} \\
        &> \frac{(\eta - \beta) + |\eta - \beta|}{2} \ge 0,
    \end{aligned}
\end{equation}
always hold for $T>0$.
To proof $\Re(\lambda_+) < 0$,
we simplify Eq.~\eqref{eq:lmd-pos} as
\begin{equation*}
\begin{aligned}
    2 \lambda_+ &= a - \frac{1}{\tau} + \sqrt{
        (a + \frac{1}{\tau})^2
        - \frac{4G^*}{\tau^2 (\eta + G^*)}
    } \\
    &=  - \frac{2}{\tau} + \frac{\beta}{\tau(\eta + G^*)} + \sqrt{
            \frac{\beta^2}{\tau^2(\eta+G^*)^2}
            - \frac{4G}{\tau^2(\eta+G^*)}
        } \\
    &= \frac{-2(\eta+G^*) + \beta}{\tau(\eta+G^*)} + \frac{
        \sqrt{\beta^2 - 4 G^*(\eta+G^*)}
        }{\tau(\eta+G^*)}
\end{aligned}
\end{equation*}
Note that if $\beta^2 - 4G^*(\eta+G^*) < 0$,
the real part
\begin{equation*}
\begin{aligned}
    \Re(\lambda_+) &= \frac{-2(\eta+G^*) + \beta}{\tau(\eta+G^*)}
    &= \frac{-2(\eta - \beta + G^*) - \beta}{\tau(\eta + G^*)}
    &< 0.
\end{aligned}
\end{equation*}

If $\beta^2 - 4G^*(\eta+G^*) > 0$,
we only need to proof
\begin{equation*}
\begin{aligned}
    -2(\eta+G^*) + \beta
    + \sqrt{\beta^2 - 4 G^*(\eta+G^*)} &< 0 \\
    \sqrt{\beta^2 - 4 G^*(\eta+G^*)} &< 2(\eta+G^*) - \beta \\
    \beta^2 - 4 G^*(\eta+G^*) &< (2(\eta+G^*) - \beta)^2 \\
    \beta^2 - 4 G^*(\eta+G^*) &< 4(\eta+G^*)^2 - 4\beta(\eta+G^*) + \beta^2 \\
    - 4 G^*(\eta+G^*) &< 4(\eta+G^*)^2 - 4\beta(\eta+G^*) \\
    - G^* &< (\eta+G^*) - \beta,
\end{aligned}
\end{equation*}
which also be satisfied.
Therefore, all eigenvalues have a negative real part,
meaning the steady-state $(\bm{R^*}, G^*)$ is an attractor.

This completes the proof.
\end{proof}

\subsection{Fixed point properties}

With $I_i = 0$,
denote $S=\sum_{i=1}^N w_i R_i, \rho_i=\frac{w_iR_i}{S}$, which satisfies $\sum_{i=1}^N \rho_i=1$, and we have
\begin{equation}
\begin{aligned}
    \tau \frac{\mathrm{d} \rho_i}{\mathrm{d} t}
    &= w_i\frac{\tau}{S} \frac{\mathrm{d}R_i}{\mathrm{d}t}-w_i\frac{\tau R_i}{S^2} \frac{\mathrm{d}S}{\mathrm{d}t} \\
    &= w_i\frac{\tau}{S}\frac{1}{\tau}\left(-R_i+\frac{\beta R_i}{\eta+G}\right)-w_i\frac{\tau R_i}{S^2}\sum_{j=1}^n w_j\frac{\mathrm{d}R_j}{\mathrm{d}t} \\
    &= w_i\frac{\tau}{S}\frac{1}{\tau}\left(-R_i+\frac{\beta R_i}{\eta+G}\right)-w_i\frac{\tau R_i}{S^2}\frac{1}{\tau}\sum_{j=1}^nw_j\left(-R_j+\frac{\beta R_j}{\eta+G}\right) \\
    &= w_i\frac{\tau}{S}\frac{1}{\tau}\left(-R_i+\frac{\beta R_i}{\eta+G}\right)-w_i\frac{\tau R_i}{S^2}\frac{1}{\tau}\left(-S+\frac{\beta S}{\eta+G}\right) \\
    &= 0
    \label{eq:propto}
\end{aligned}
\end{equation}

With the inputs exerted to the system, we have
\begin{equation}
    \begin{aligned}
    \tau \frac{\mathrm{d}\rho_i}{\mathrm{d} t} &= w_i\frac{\tau}{S} \frac{\mathrm{d}R_i}{\mathrm{d}t}-w_i\frac{\tau R_i}{S^2} \frac{\mathrm{d}S}{\mathrm{d}t} \\
    &=w_i\frac{\tau}{S}\frac{1}{\tau}\left(-R_i+\frac{I_i+\beta R_i}{\eta+G}\right)-w_i\frac{\tau R_i}{S^2}\sum_{j=1}^n w_j\frac{\mathrm{d}R_j}{\mathrm{d}t} \\
    &=w_i\frac{\tau}{S}\frac{1}{\tau}\left(-R_i+\frac{I_i+\beta R_i}{\eta+G}\right)-w_i\frac{\tau R_i}{S^2}\frac{1}{\tau}\sum_{j=1}^nw_j\left(-R_j+\frac{I_j+\beta R_j}{\eta+G}\right) \\
    &=w_i\frac{1}{S}\left(-R_i+\frac{I_i+\beta R_i}{\eta+G}\right)-w_i\frac{R_i}{S^2}\left(-S+\frac{\beta S}{\eta+G}+\sum_{j=1}^n\frac{w_jI_j}{\eta+G}\right) \\
    &=w_i\left(\frac{I_i}{S(\eta+G)}-\frac{R_i}{S^2(\eta+G)} \sum_{j=1}^nw_jI_j\right)
    \end{aligned}
    \label{eq:propto1}
\end{equation}

When the system has converged to the fixed point, we have $I_i=R_i^*(\eta+G^*-\beta)$, and
\begin{equation}
\begin{aligned}
\tau \frac{\mathrm{d}\rho_i}{\mathrm{d} t} &= w_i \left( \frac{I_i}{S(\eta+G^*)}-\frac{R^*_i}{S^2(\eta+G^*)} \sum_{j=1}^nw_jI_j \right) \\
&= w_i \left( \frac{R_i^*(\eta+G^*-\beta)}{S(\eta+G^*)}-\frac{R_i^*(\eta+G^*-\beta)\sum_{j=1}^nw_jR_j^*}{S^2(\eta+G^*)} \right) \\
&= w_i \left(\frac{R_i^*(\eta+G^*-\beta)}{S(\eta+G^*)}-\frac{R_i^*(\eta+G^*-\beta)S}{S^2(\eta+G^*)} \right) \\
&= 0
\end{aligned}
    \label{eq:propto2}
\end{equation}

It means that with inputs exerted to the system, the balances among proportions $\rho_i$ will initially be broken, in other words, the previous *memory* of the system will be *destroyed*, and $\rho_i$ will return to constant when the system converges to a new fixed point, forming a new representation of inputs.
% \begin{equation}
%     \begin{aligned}
%     \tau \frac{\mathrm{d} \rho_i}{\mathrm{d} t}
%     &= \frac{\tau}{S} \frac{\mathrm{d}R_i}{\mathrm{d}t}
%         - \frac{\tau R_i}{S^2} \frac{\mathrm{d}S}{\mathrm{d}t} \\
%     &= \frac{\tau}{S}\frac{1}{\tau}\left(-R_i+\frac{\beta R_i}{\eta+G}\right)
%         - \frac{\tau R_i}{S^2}\sum_{j=1}^N \frac{\mathrm{d}R_j}{\mathrm{d}t} \\
%     &= \frac{1}{S}\left(-R_i+\frac{\beta R_i}{\eta+G}\right)
%         - \frac{\tau R_i}{S^2}\frac{1}{\tau}\sum_{j=1}^N\left(
%             -R_j+\frac{\beta R_j}{\eta+G}
%         \right) \\
%     &= \frac{1}{S}\left(-R_i+\frac{\beta R_i}{\eta+G}\right)
%         - \frac{R_i}{S^2}\left(
%             -S+\frac{\beta S}{\eta+G}
%         \right) \\
%     &= 0
%     \label{eq:propto}
%     \end{aligned}
% \end{equation}

%%%%%%%%%%%%%%%%%%%%%%%%%%%%%%%%%%%%%%%%%%%%%%%%%%%%%%%%%%%%
\end{document}